\documentclass[twoside,twocolumn,english,aps,pra,superscriptaddress,showpacs]{revtex4}
\usepackage[T1]{fontenc}
\usepackage[latin1]{inputenc}
\usepackage{amssymb}

\makeatletter

\usepackage{graphicx}
\usepackage{babel}
\makeatother
\begin{document}
\newcommand {\ve}[1]{\mbox{\boldmath \(#1\)}}\newcommand {\ga}{\raisebox{-.5ex}{\(\stackrel{>}{\sim}\)}}
\newcommand {\la}{\raisebox{-.5ex}{\(\stackrel{<}{\sim}\)}}
\newcommand{\beq}{\begin{equation}}
\newcommand{\eeq}{\end{equation}}
\newcommand{\bea}{\begin{eqnarray}}
\newcommand{\eea}{\end{eqnarray}}

\title{Static properties of positive ions in atomic Bose-Einstein condensates}

\author{P. Massignan}

\affiliation{\O rsted Laboratory, H.\ C.\ \O rsted Institute, Universitetsparken
5, DK-2100 Copenhagen \O, Denmark}

\author{C. J. Pethick}

\affiliation{NORDITA, Blegdamsvej 17, DK-2100 Copenhagen \O, Denmark}

\author{H. Smith}

\affiliation{\O rsted Laboratory, H.\ C.\ \O rsted Institute, Universitetsparken
5, DK-2100 Copenhagen \O, Denmark}

\date{\today{}}

\begin{abstract}
The excess number of atoms around an ion immersed in a Bose-Einstein
condensate is determined as a function of the condensate density far
from the ion. We use thermodynamic
arguments to demonstrate that in the limit of low densities the excess number of atoms is proportional to
the ratio of the atom-ion and atom-atom scattering lengths. For
denser systems we calculate the excess number from solutions of the
Gross-Pitaevskii equation using a model potential that has a $1/r^{4}$
attraction coming from the polarization of the neutral atoms and a
hard core repulsion at short distances. We show that there
exist in general many solutions to the
Gross-Pitaevskii equation for a given condensate density, the maximum number
of solutions being related to the number of bound states of the
Schr\"odinger equation for the same potential. With increasing density,
pairs of these solutions merge and disappear, implying a discontinuous
change of the state of the system.

\end{abstract}


\maketitle

\section{Introduction}
For the helium liquids, measurements on ions have served as a valuable
probe of liquid properties. As examples we may mention the use of ions
to put in evidence vortex lattices in rotating liquid $^{4}$He and
the measurement of ionic mobilities in superfluid $^{3}$He to elucidate
scattering processes. The first
experiments on ions in an ultracold gas of $^{87}$Rb atoms were reported
by the group in Pisa \cite{Pisa}, who produced ions by irradiating
a rubidium condensate with laser pulses which ionize atoms through
one- and two-photon absorption processes.

Theoretically, the capture of atoms into weakly bound states of the
atom-ion potential has been considered in Ref.\ \cite{Cote'}. In
this paper we consider the structure of a positive ion in a Bose-Einstein
condensate when there is no capture of atoms into bound states, and
in particular we calculate the excess number of atoms associated with
an ion. We shall demonstrate that this number is typically of order 10$^{2}$,
 either positive or negative.

The interaction between an atom and a positively charged alkali-metal
ion (charge $e$), which are separated by a distance $r$, is given
at large distances by the polarization potential caused by the electrostatic
field $\mathcal{E}_{\rm es}$ due to the ion, ${\mathcal{E}}_{\rm es}=e/4\pi\epsilon_{0}r^{2}$, which
 gives rise to a change in the energy of the neutral atom 
given by $V=-\alpha{\mathcal{E}}^{2}/2$,
where $\alpha$ is the polarizability of the atom. Writing the polarizability
as $\alpha=4\pi\epsilon_{0}\tilde{\alpha}$, where $\tilde{\alpha}$
has the dimension of volume, the energy shift of an atom due to the
ion becomes 
\begin{equation}
V(r)=-\tilde{\alpha}\frac{e_{0}^{2}}{2r^{4}},
\label{pot}
\end{equation}
 where $e_{0}^{2}\equiv e^{2}/4\pi\epsilon_{0}$. At short distances
($r\la10a_{0})$ the potential has a repulsive core. An important
characteristic length, which we denote by $\beta_{4}$, may be
identified by equating the kinetic energy $\hbar^{2}/2m\beta_{4}^{2}$
to the potential energy $V(\beta_{4})$, resulting in
\begin{equation}
\beta_{4}=\sqrt{\frac{\tilde{\alpha}}{a_{0}}\frac{m}{m_{\rm e}}}.
\label{beta4}
\end{equation}
 Here $m$ denotes the mass of a neutral atom, $m_{\rm e}$ is the
electron mass and $a_{0}\equiv\hbar^{2}/m_{\rm e}e_{0}^{2}\sim0.53$ \AA \
is the Bohr radius. Using the measured values $\tilde{\alpha}=320\, a_{0}^{3}$
for $^{87}$Rb and $\tilde{\alpha}=163\, a_{0}^{3}$ for $^{23}$Na, one finds $\beta_{4}^{\rm Rb}\approx 7150 a_0$ and $\beta_{4}^{\rm Na}\approx 2620a_0$. The quantity $\beta_{4}$
gives the distance from the ion beyond which the zero-energy atom-ion wave function ceases to
oscillate, and it sets the scale of atom-ion scattering lengths, but their actual values
depend on the details of the potential at short distances.

We begin by deriving from thermodynamics a general expression for the excess number of atoms around an ion and show
that in
dilute systems the excess number depends only on the ratio of the atom-ion and atom-atom scattering
lengths. As we shall see, this approach suggests that the number of
atoms associated with an ion is typically of the order of 10--100,
but that it may be either positive or negative.
In denser systems the
excess number must be obtained from microscopic considerations, and we
shall determine the structure of an ion immersed in a Bose-Einstein
condensate at zero temperature, assuming that atom-atom interactions
may be described within the framework of the Gross-Pitaevskii (GP)
mean-field approach. We present solutions of the GP equation for a
number of potentials which include a hard core repulsion, an
attractive square well, and one which resembles the atom-ion
interaction, a hard core with a $1/r^{4}$ attraction at larger
distances.

For a given inner boundary condition, the Schr\"odinger equation has
only one solution for a given value of the energy.  By contrast, the
GP equation, because it is nonlinear, can have more than one solution
for a given chemical potential.  For potentials like the atom-ion one
that can support two-body bound states, we shall find that at low
densities there are $2\nu_{\rm S}+1$ solutions of the GP equation,
where $\nu_{\rm S}$ is the number of bound states of the Schr\"odinger
equation for the same potential.  With increasing density, pairs of
solutions merge and disappear until there is only a single solution
with no nodes. We shall illustrate this behavior for two potentials,
an attractive square well and one with an attractive $1/r^4$ tail.  An
important question is which of these solutions is physically relevant.
At low condensate densities, one expects the wave function close to
the ion to resemble the zero-energy solution of the Schr\"odinger
equation, and to have $\nu_{\rm S}$ nodes. This will be the case
unless inelastic processes can populate lower-lying states.  We find
that with increasing condensate density, this solution ceases to
exist. This indicates that the evolution of the state with density
cannot be continuous even in the absence of inelastic processes.

The plan of the paper is as follows.  In Sec.\ \ref{sec:thermo} we
present thermodynamic considerations.  Section \ref{microscopic}
contains a description of the asymptotic behavior of the condensate
wave function far from the ion.  In Sec.\ \ref{sec:model-pot} we
consider two simple potentials to illustrate important general
features of our results, and in Sec.\ \ref{polpot} we analyse the case
of a potential that, like the actual atom-ion potential, behaves as
$r^{-4}$ at large distances.  We calculate the excess number of atoms
from numerical solutions of the Gross-Pitaevskii equation for a given
background condensate density. The concluding section, Sec.\ \ref{concl},
discusses our main results. In an appendix, we address the question
of the validity of the Gross-Pitaevskii equation in the present
context.

\section{Thermodynamic considerations}

\label{sec:thermo}

We wish to calculate the excess number of particles associated with
an ion. To define this quantity precisely, we imagine adding an ion
to a condensate. This will generally change the density of atoms far
from the ion by an amount that varies as $1/V$, where $V$ is the
volume of the system. A natural definition of the excess number of
atoms $\Delta N$ associated with the ion is the number of particles that
must be added to keep the atom chemical potential constant, since
this will ensure that the properties of the condensate far from the
ion are unaltered by the addition of the ion. 
In terms of the microscopic density of atoms $n(r)$ around the ion, the excess number is given by 
\begin{equation}
\Delta N=4\pi\int_0^{\infty}dr\ r^{2}\left[n(r)-n_{0}\right],
\label{DeltaN} 
\end{equation}
where $n_0$ is the density of atoms at large distances from the ion.

This is analogous to
what has been done earlier to calculate the excess number of $^{4}$He
atoms associated with a $^{3}$He impurity in liquid $^{4}$He \cite{bbp}.
We shall denote the energy per unit volume as ${\cal E}(n_\mathrm{a},n_\mathrm{i})$,
where $n_\mathrm{a}$ and $n_\mathrm{i}$ are the number densities
of atoms and ions, respectively. The chemical potential of the atoms
is given by
\begin{equation}
\mu_\mathrm{a}=\frac{\partial {\cal E}}{\partial n_\mathrm{a}},
\end{equation}
 and therefore the condition that this be unchanged by adding one
ion and $\Delta N$ atoms is
\begin{equation}
\frac{\partial^{2}{\cal E}}{\partial n_\mathrm{a}\partial n_\mathrm{i}}
+\frac{\partial^{2}{\cal E}}{\partial n_\mathrm{a}^2}\Delta N=0,
\end{equation}
 or
 \begin{equation}
\Delta N=-{\frac{\partial^{2}{\cal E}}{\partial n_\mathrm{a}\partial n_\mathrm{i}}}\left/\frac{\partial^{2}{\cal E}}{\partial n_\mathrm{a}^2}\right. .\label{1stDeltaN}
\end{equation}
 When the density of ions is sufficiently low, $\partial {\cal E}/\partial n_\mathrm{i}$
is equal to the energy change $\epsilon_\mathrm{i}$ when one ion
is added to the condensate, and therefore \begin{equation}
\Delta N=-{\frac{\partial\epsilon_\mathrm{i}}{\partial n_\mathrm{a}}}
\left/\frac{\partial^{2}{\cal E}}{\partial n_\mathrm{a}^2}\right. .
\end{equation}

One may also calculate $\Delta N$ from the change $\Delta F$ in the
thermodynamic potential $F=E-\mu_\mathrm{a}N$ when
a single ion is added to the system at constant $\mu_\mathrm{a}$. Here $E$ is the total energy and $N$ the total number of atoms.
Since the number of atoms is given by
\begin{equation}
N=-\frac{\partial F}{\partial\mu_\mathrm{a}},
\end{equation}
it follows immediately that
\begin{equation}
\Delta N=-\frac{\partial\Delta F}{\partial\mu_\mathrm{a}}.
\label{DeltaN-FreeEnergy}
\end{equation}
Provided the volume considered
is large compared with the scale of the atom excess around the ion,
$\Delta F$ will be independent of the volume.

Let us begin by making estimates for a dilute gas.  Provided the scattering of atoms by atoms and of atoms by ions may be treated as independent binary events, the energy density may be expressed in terms of the
scattering lengths associated with the atom-atom and atom-ion
interactions.  If ion-ion interactions are neglected,
we may write 
\begin{equation}
{\cal E}(n_\mathrm{a},n_\mathrm{i})=
\frac{1}{2}U_\mathrm{aa} n_\mathrm{a}^{2}+U_\mathrm{ai}n_\mathrm{a} n_\mathrm{i},
\end{equation}
and therefore from Eq.\ (\ref{1stDeltaN}) we obtain
\begin{equation}
\Delta N=-\frac{U_\mathrm{ai}}{U_\mathrm{aa}}.
\end{equation}
The mean-field interaction constant $U_{jl}$ for species $j$ and $l$,
which may be either atoms (a) or ions (i), is related to the
scattering length $a_{jl}$ by
\beq
U_{jl}=\frac{2\pi\hbar^{2}a_{jl}}{m_{jl}}, 
\label{U}
\eeq
where
$m_{jl}=m_j m_l/(m_j+m_l)$
is the reduced mass of the two particles. Our result can therefore be expressed as
\begin{equation}
\Delta N=-\frac{m_\mathrm{aa}}{m_\mathrm{ai}}\frac{a_\mathrm{ai}}{a_\mathrm{aa}}.
\label{eq: deficit in the dilute limit1}
\end{equation}
If, as in Ref.\ \cite{Pisa}, the ion is obtained by photoionization
of the condensate itself, the latter expression reduces to
\begin{equation}
\Delta N=-a_\mathrm{ai}/a_\mathrm{aa}.
\label{DeltaNscatlength}
\end{equation}
To obtain an order of magnitude estimate of the excess number of atoms
associated with an ion, we note that the characteristic scale for the
magnitudes of atom-ion scattering lengths
$\left|a_{\mathrm{ai}}\right|$ is set by $\beta_{4}$, given in Eq.\
(\ref{beta4}), while the scale for the magnitudes of atom--atom scattering
lengths $\left|a_{\mathrm{aa}}\right|$ is set by
\begin{equation}
\beta_6= \left(2\frac{m}{m_{\rm e}}C_6 \right)^{1/4}  a_0.
\label{beta6}
\end{equation}
Here $C_6$ is the coefficient of $r^{-6}$ in the van der Waals
interaction, expressed in atomic units.
Thus we arrive at the estimate
\begin{equation}
|\Delta N|\sim \frac{\beta_4}{\beta_6} \sim
\left(\frac{m}{2m_{\rm e}} \frac{{\tilde \alpha}^2}{C_6}\right)^{1/4},
\label{DeltaN_thermo}
\end{equation}
which is of order 35 for Rb and 25 for Na.

The fact that the excess number of atoms is so large indicates that it
may well be a poor approximation to regard the ion as a free particle,
with mass equal to the bare ion mass.  Rather, the recoil of the ion
will be suppressed by the other atoms surrounding the ion, and if
$\Delta N \gg 1$ it will be a better approximation to regard the ion
as being stationary.  In that case the excess number of atoms
will be given by
\beq
\Delta N= - \frac{a_{\rm ai}(m)}{2 a_{\rm aa}},
\label{DeltaNinfmass}
\eeq
 where the argument of $a_{\rm ai}$ indicates that the scattering
 length is to be evaluated for a reduced mass equal to the atom mass.
 Expression (\ref{DeltaNinfmass}) gives a value for $\Delta N$ that is
 typically of the same order of magnitude as that given by Eq.\
 (\ref{DeltaN_thermo}).  However, we stress the fact that the estimate
 for $\Delta N$ depends sensitively on the value of the effective mass
 of the ion, since the atom-ion potential has many bound states, and
 therefore relatively small changes in the reduced mass can result in
 large changes in the scattering length.  Given that in the limit of
 low atom density the magnitude of the excess number of atoms is
 expected to be very much greater than unity, the result
 (\ref{DeltaNscatlength}) will generally not give a realistic estimate
 even in that case.

The perturbation induced by the ionic potential is very
strong. Therefore the question arises of whether the customary
assumption of an essentially zero range for the atom-atom interaction
is valid.  We address this point in Appendix A, where we argue that
the corrections to the GP result should not be large for the
properties of interest here.

\section{Microscopic theory}

\label{microscopic}

We now turn to microscopic considerations.  Since, as we shall see,
the distortion of the condensate wave function in the vicinity of an
ion extends to large distances from the ion and involves many atoms,
we expect that the effective mass of an ion and its dressing cloud
will be much larger than that of an atom, and we may regard the ion as
being static.  To describe the structure of the condensate in the
vicinity of an ion we must therefore calculate the structure of the
condensate in a static external potential given by the atom-ion
interaction.  Provided the length scale on which the condensate wave
function $\psi$ varies in space is sufficiently large, we may do this
by employing the Gross-Pitaevskii equation with the interaction
of atoms with the ion included as an external potential,
\begin{equation}
\left[-\frac{\hbar^{2}}{2m}\nabla^{2}+V(r)+U_{0}\left|\psi\right|^{2}\right]\psi=\mu\psi.
\label{gp}
\end{equation}
 Here and in what follows we shall denote the chemical potential of an atom by $\mu$, and for simplicity we
have written
$U_{0}\equiv U_{\mathrm{aa}} =4\pi\hbar^{2}a_{\mathrm{aa}}/m$,
since $m_{\rm aa}=m/2$. We wish to find
solutions that tend to a constant at large
distances from the ion, and since the potential is
spherically symmetric, these
solutions depend only on the radial coordinate $r$.
Thus Eq.\ (\ref{gp}) becomes

\begin{equation}
\left[-\frac{\hbar^{2}}{2m}\frac{d^{2}}{d r^2}+V(r)+U_{0}\frac{\left|\chi\right|^{2}}{r^2}\right]\chi=\mu\chi,
\label{gpradial}
\end{equation}
where $\chi=r\psi$.

The behavior of the condensate wave function at large distances depends
on the nature of the potential $V(r)$.
On linearizing the GP equation (\ref{gp}) and making use of the fact that
the chemical potential is related to the condensate wave function
$\psi_0$ at large distances by the relation $\mu = n_0 U_0$ where
$n_0=|\psi_0|^2$, one finds that
the deviation
\begin{equation}
\delta \psi =\psi -\psi_0
\label{deltapsi}
\end{equation}
 of the condensate wave
function
from its asymptotic value satisfies the
linearized GP equation

\begin{equation}
\left(-\frac{\hbar^{2}}{2m}\nabla^{2}+V(r) + 2U_0 n_0\right)\delta \psi=-V(r)\psi_0.
\label{linearGP}
\end{equation}

For potentials with a finite range, one may neglect the potential at
large distances from the ion, and the deviation
that vanishes for $r\rightarrow \infty$ is thus given by
\begin{equation}
\delta \psi \propto \frac{{\rm e}^{-k_\xi r}}{r},
\label{deltapsir}
\end{equation}
where $k_\xi=\sqrt 2/\xi$ and $\xi$ is the healing length for
the bulk condensate,
\begin{equation}
\xi=\frac{1}{\sqrt{8\pi a_{\mathrm{aa}}n_{0}}}.
\label{healing}
\end{equation}

For a potential, such as the atom-ion potential, that falls off at large distances less rapidly than the
solution (\ref{deltapsir}),
the behavior is different. The leading term
in the solution for large
$r$ is then
the Thomas--Fermi result $\psi_{\rm TF}$, given by
\begin{equation}
V(r)+U_0|\psi_{\rm TF}|^2=\mu,
\end{equation}
which, for the
atom-ion potential with asymptotic form given by
Eq.\ (\ref{pot}), amounts to
\begin{equation}
n_{\mathrm{TF}}(r)=n_{0}-\frac{V(r)}{U_{0}}=
n_{0}\left(1+\frac{(\xi\beta_{4})^{2}}{r^{4}}\right)\label{TFsolution}
\end{equation}
or, to first order in $V$,

\begin{equation}
\psi_{\rm TF}\approx \psi_0 - \frac{V(r)}{2U_0 \psi_0},
\label{deltapsiTF}
\end{equation}
where we have taken $\psi_0$ to be real. The density perturbation at large distances is seen to be
always positive. Corrections to
this result for smaller $r$ may be calculated from Eq.\
(\ref{linearGP}) by neglecting the potential on the left hand side
of the equation. The resulting differential equation may be
written in terms of a function $\delta\chi$ defined by
$\delta\chi=r\delta\psi$,
\begin{equation}
\left(-\frac{d^2}{dr^2}+k_{\xi}^2\right)\delta\chi=-\frac{2m}{\hbar^2}rV(r)\psi_0.
\label{chieq}
\end{equation}
This differential equation may be solved exactly in terms of
exponential integrals, the two linearly independent solutions of the
homogeneous equation being $\exp(-k_{\xi}r)$ and $\exp(k_{\xi}r)$. By
inspection of (\ref{chieq}) it is evident that the leading term for
large $r$ of the particular solution to the inhomogeneous equation is
given by $\delta \chi=- (2m/\hbar^2k_{\xi}^2)rV(r)\psi_0$, which
yields the Thomas-Fermi expression (\ref{deltapsiTF}). The correction
to this result may be obtained from the exact solution, but it is
simpler to iterate Eq.\ (\ref{chieq}) by moving the term
$d^2\delta\chi/dr^2$ to the right hand side and replacing $\chi$ in it
by the Thomas-Fermi solution.  This results in
\begin{equation}
\frac{\delta \psi}{\psi_0}=-\frac{2m}{\hbar^2k_{\xi}^2}V(r)\left(1+\frac{12}{k_{\xi}^2r^2}\right).
\label{chieq1}
\end{equation}
The leading correction to the Thomas-Fermi result for $\delta\psi$
given in (\ref{deltapsiTF}) is thus seen to be proportional to
$r^{-6}$. Since we have already neglected the potential energy on
the left hand side of (\ref{linearGP}), we cannot by this method
obtain higher-order corrections to the particular solution than
the one exhibited in (\ref{chieq1}).

By keeping in the general solution only the exponentially decaying
term we thus get the asymptotic result
\begin{equation}
\frac{\delta \psi}{\psi_0} \sim  -\frac{V(r)}{2n_0U_0 }\left(1+\frac{6\xi^2}{r^2}\right)
+C\frac{{\rm e}^{-k_\xi r}}{r}, \label{deltapsilarger}
\end{equation}
where $C$ is an arbitrary constant.

For $r \rightarrow \infty$ the asymptotic behavior of the solution is
always given by the TF result.  However, whether or not this behavior
is relevant for determining the structure of most of the cloud of
atoms surrounding the ion depends on the relative size of the two
characteristic lengths, $\beta_4$ and $\xi$. On the one hand, for
$\beta_4\gg\xi$ most of the cloud will be described by the TF
approxmation, and only at distances less than $\sim \xi$ will the
exponential term become important.  On the other hand, for
$\xi\gg\beta_4$ (i.e.\ for low external density) the structure will be
dominated by the exponential term, and the TF tail will become
important quantitatively only at very large $r$.  At shorter distances
from the ion, the mean-field energy becomes small compared with the
atom-ion potential and the GP equation reduces to a good
approximation to the Schr\"odinger equation.

\section{Simple model potentials}
\label{sec:model-pot}

Before presenting results for the attractive $1/r^{4}$
potential we begin by examining two simpler model
potentials, a repulsive hard-core and a spherical well.

\subsection{Hard-core potential} 
Consider an interacting Bose-Einstein condensed gas in the
presence of a repulsive hard-core potential of radius $R$.
This model may be treated analytically in both the small and large core radius limits.
The solution to the GP equation at large distances from the ion is given by Eq.\ (\ref{deltapsir}),
\begin{equation}
\psi\simeq\sqrt{n_{0}}\left(1+C\frac{\exp(-k_\xi r)}{r}\right).
\label{Yuk}
\end{equation}
If one assumes that this expression holds for all $r$ greater than
$R$, we can determine the constant of proportionality $C$ by imposing
the boundary condition $\psi(R)=0$. This gives $C=-R{\rm
e}^{k_{\xi}R}\sqrt{n_{0}}$.  For $r$ close to $R$ this has the
form of the scattering solution for the Schr\"odinger equation,
$\psi=1-R/r$.  In fact, for $R/\xi\ll 1$ this solution becomes
essentially exact, since this function fails to satisfy the GP
equation only in the region where $r \simeq R$, and in this region the
total change in the slope $d\chi/dr$ of the radial wave function is
small and may be neglected.  As an illustration of this fact, we
calculate the excess number of particles, which is given by Eq.\
(\ref{DeltaN}), and find

\bea
\Delta N\!&\!=\!&\!4\pi n_0 \int_{R}^{\infty}\!\!\!dr\, r^{2}\left(-2R\frac{e^{-k_{\xi}(r-R)}}{r} + R^2\frac{e^{-2k_{\xi}(r-R)}}{r^2}\right) \nonumber\\
\!&\!\!&\ \ \ \ \ \ \ \ \ \ \ \ \ \ \ \ \ \ \ \ \ \ \ \ \ \ \ \ \ \ \ \ \ \ \ \ \ \ \ \ \ \ \ \ \ \ \ 
 -\frac {4\pi n_0 R^3}{3}=\nonumber\\
\!&\!=\!&\!-4\pi n_0\left( R\xi^2+ \frac{3R^2 \xi}{2\sqrt 2}+\frac{R^3}{3} \right).
\eea
For $\xi \gg R$, this reduces to 
\beq
\Delta N=-\frac{R}{2a_{\rm aa}}.
\eeq
 Let us now compare this result with the one derived on the basis of thermodynamic arguments. 
For a hard-core potential the scattering length coincides with
the core radius. Since we have assumed the ion to be stationary, its effective mass is taken to be infinitely large, and therefore the reduced mass for the ion and an atom is $m$, rather than the value $m/2$ one obtains for an ion and an atom with equal masses. Thus, this
result is in precise agreement with  Eq.\ (\ref{eq: deficit in the dilute limit1}).

When the core radius is much larger than the healing length,
the wave function reaches its asymptotic value on a length scale that
is short compared to $R$. In Eq.\ (\ref{gpradial}) we can therefore
 replace the factor $1/r^{2}$ appearing in the nonlinear
term by the constant $1/R^{2}$ and we are left with an effectively
one-dimensional GP equation whose solution is
\begin{equation}
\psi=\sqrt{n_{0}}\tanh\frac{r-R}{\sqrt{2}\xi}, \;\; r\geq R,
\label{eq: 1Dsoliton}
\end{equation}
and zero otherwise, as may be seen by inspection.
The excess number of particles is given by
\beq
\Delta N= -\frac{4}{3}\pi R^{3}n_{0} -\frac{R^{2}}{\sqrt{2}a_{\mathrm{aa}}\xi},
\eeq
where the leading term is due to exclusion of atoms from the core.

\subsection{Attractive square well}

We next consider a more physical potential, an attractive well:
\beq
V(r)=-\frac{\hbar^{2}k_{0}^{2}}{2m}, \;\;\;r<R, 
\eeq
$V(r)=0$ otherwise. Like the actual ion-atom potential, this can have
bound states for the two-body problem. With this potential we shall be
able to examine how solutions of the GP equation disappear as the
condensate density increases. The GP equation (\ref{gpradial}) reads
\begin{equation}
\left[-\frac{1}{2}\frac{d^{2}}{d r^{2}}-\frac{k_{0}^{2}}{2}\theta(R-r)+4\pi a_{\rm aa}\left(\frac{\left|\chi\right|^{2}}{r^{2}}-n_0\right)\right]\chi=0
\end{equation}
 (where $\chi=r\psi$ and $\theta(x)$ is the step function)
and the scattering length for this potential is\begin{equation}
a=R\left(1-\frac{\tan k_0 R}{k_0R}   \right).\end{equation}

Since this equation is nonlinear, there can be multiple solutions for
the same boundary conditions (i.e.\ the same bulk density $n_{0}$). As
we will show in the following, for small $n_0$ it has $2\nu_{\rm S}+1$
solutions, where $\nu_{\rm S}$ is the number of nodes of the zero-energy
Schr\"{o}dinger solution $\psi_{\rm S}$ or, equivalently, the number of bound states
of the Schr\"odinger equation.  In the low background density limit,
inside the well the solution with the maximum number of nodes approaches
$\psi_S$, i.e.\ $\psi(r)\propto\sin(k_{0}r)/r$, while outside it
tends towards the uniform density $n_0$ with the asymptotic behavior
given in Eq.\ (\ref{deltapsir}), $\delta\psi(r)\propto\exp(-k_{\xi}r)/r$.

With increasing $n_{0}$, the mean-field repulsion between the atoms
makes the effective potential shallower, which tends to push nodes of
the wave function outwards.  At the same time, the increase in the
chemical potential has the opposite effect on the nodes. What we find
is that if the zero-energy solution of the Schr\"odinger equation has
$\nu_{\rm S}$ nodes, for low condensate densities the GP equation has
one solution with no nodes, and {\it two} solutions with any nonzero 
number of
nodes less than or equal to $\nu_{\rm S}$.

To demonstrate this, we analyze separately the behavior of the wave
function inside and outside the well, and match them at some
intermediate point, which for this particular potential we take to be
the edge of the well. Specifically, we integrate out from the origin,
where $\chi=0$, for different choices of the derivative of $\chi$ at
$r=0$ and calculate $\psi$ and $\psi'$ at the boundary $r=R$. These
trace a curve in $\psi -\psi'$ space. Then we integrate inwards from
large distances, where the solution is defined by the proportionality
constant $C$ of the Yukawa asymptotic form, Eq.\ (\ref{deltapsir}).
As $C$ is varied, another curve in $\psi -\psi'$ space is traced out.
If the mean-field interaction could be neglected for $r<R$, the ratio
$\psi'(R)/\psi(R)$ would not depend on the normalization of the
wavefunction, and therefore the curve corresponding to the inner
boundary would be a straight line through the origin.  In the presence
of atom-atom interactions, the ratio $\psi'/\psi$ obtained by
integrating outwards traces out a spiral.  For low $n_0$ this crosses
the $\psi'$ axis a number of times equal to the number of nodes the
zero-energy solution of the Schr\"odinger equation has inside the
potential.  This follows from the observation that for low $\chi'(0)$
the solution will have the same number of nodes inside the potential
as the zero-energy solution of the Schr\"odinger equation, while for
very large values of $\chi'(0)$ the effects of the mean field will be
so strong that the solution has no nodes inside the potential.

\begin{figure}
\includegraphics[width=\columnwidth,angle=0,clip=]{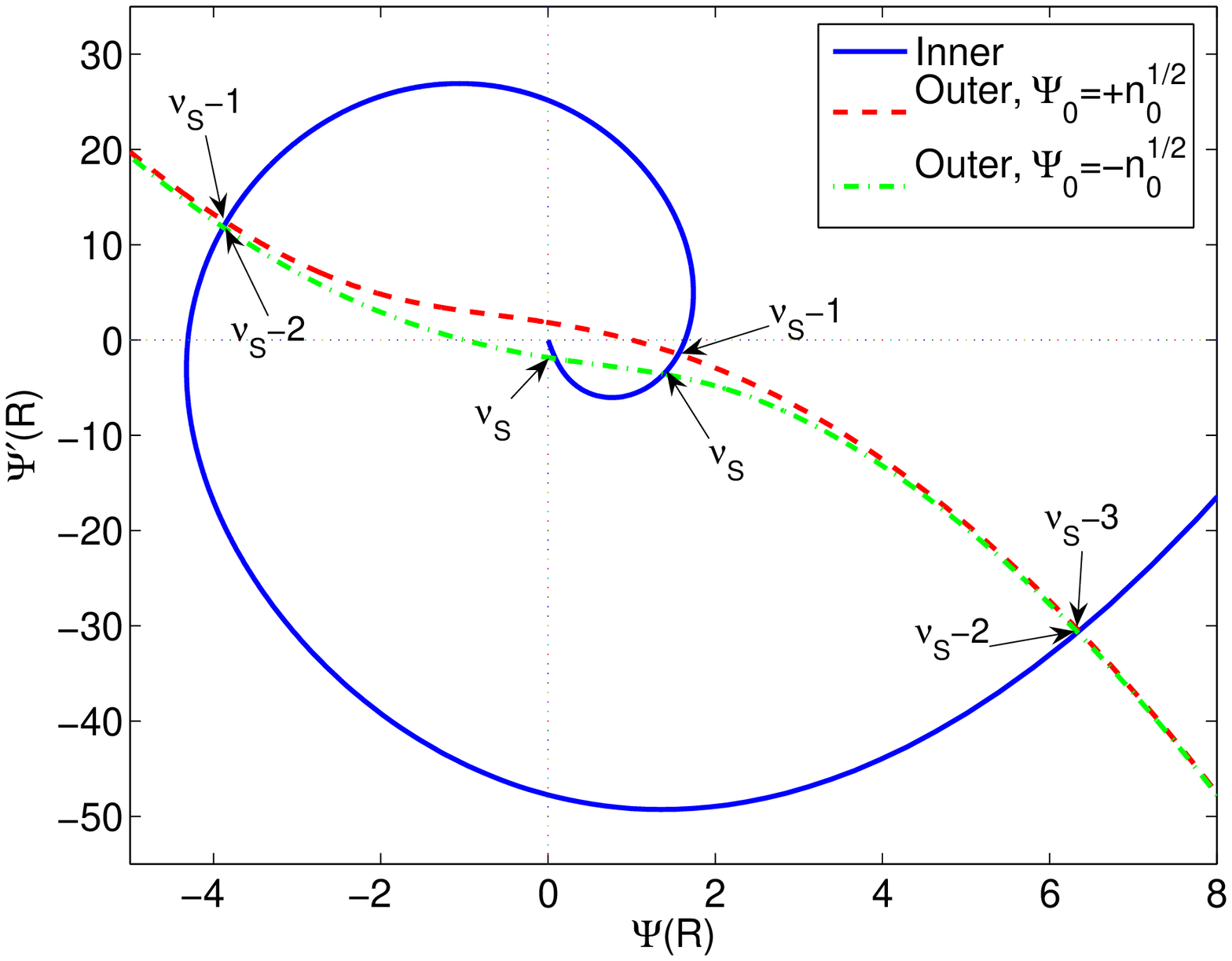}
\includegraphics[width=\columnwidth,angle=0,clip=]{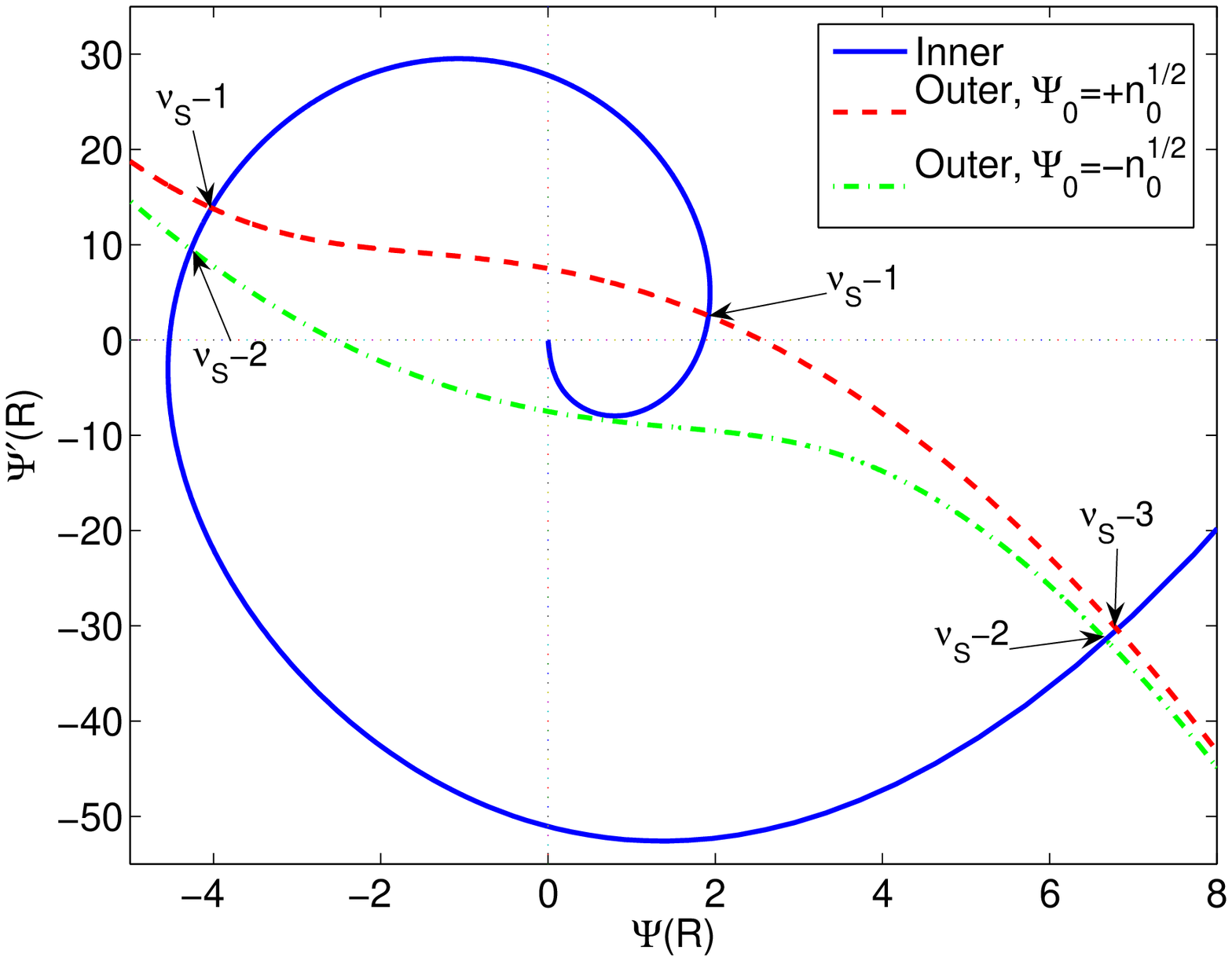}

\caption{\label{cap:spirals}(Color online) Behavior of $\psi(R)$ and $\psi'(R)$ 
  for the solution inside the well (solid line) and outside it (dashed
  and dot-dashed lines for $\psi (r\rightarrow \infty)=\pm \psi_0$,
  respectively). In the plots we have set $k_0 R= 9$, which gives
  three bound states for the Schr\"odinger equation.  We measure
  energies in units of $\hbar^2 k_0^2/2m$ and lengths in units of $R$.
  The calculations were performed for $U_0=0.45$ in these units, but
  results for other values of $U_0$ may be obtained by scaling, since
  for a given chemical potential, $\psi$ and $\psi'$ vary as
  $U_0^{-1/2}$.  The symbols near intersections indicate the number of
  nodes of the solution, and for this case $\nu_{\rm S}=3$. The upper
  panel (a) is for $\mu=0.45$, and the lower one (b) for $\mu=2.9$,
  just above the value $\mu=2.52$ at which the two solutions with 3
  nodes merge and disappear.}
\end{figure}

The corresponding plot obtained by integrating inwards has two
branches, depending on whether $\psi(r\rightarrow \infty)$ is positive
or negative. Examples of the plots are given in Fig.\ 
\ref{cap:spirals} for parameters such that $\nu_{\rm S}=3 $. For low
$n_0$, there are $2\nu_{\rm S}+1$ intersections of the two sets of
curves, corresponding to solutions of the GP equation.  This is
illustrated in Fig.\ \ref{cap:spirals}a.  As $n_0$ increases, pairs of
solutions with the same number of nodes merge and disappear, as shown
in Fig.\ \ref{cap:spirals}b.  Eventually, at sufficiently high values
of $n_0$ only the nodeless solution survives.

\begin{figure}
\includegraphics[width=\columnwidth,clip=]{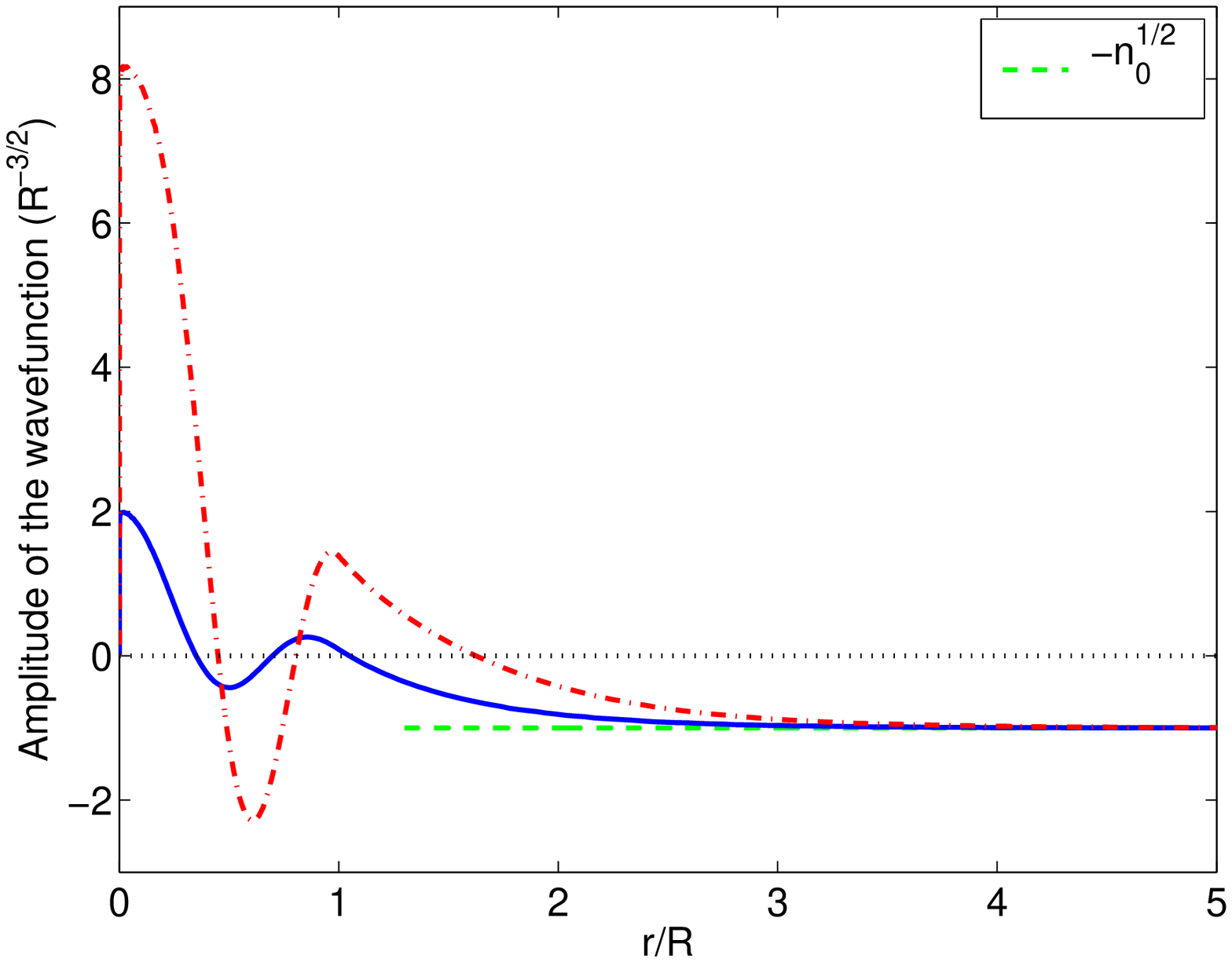}
\includegraphics[width=\columnwidth,clip=]{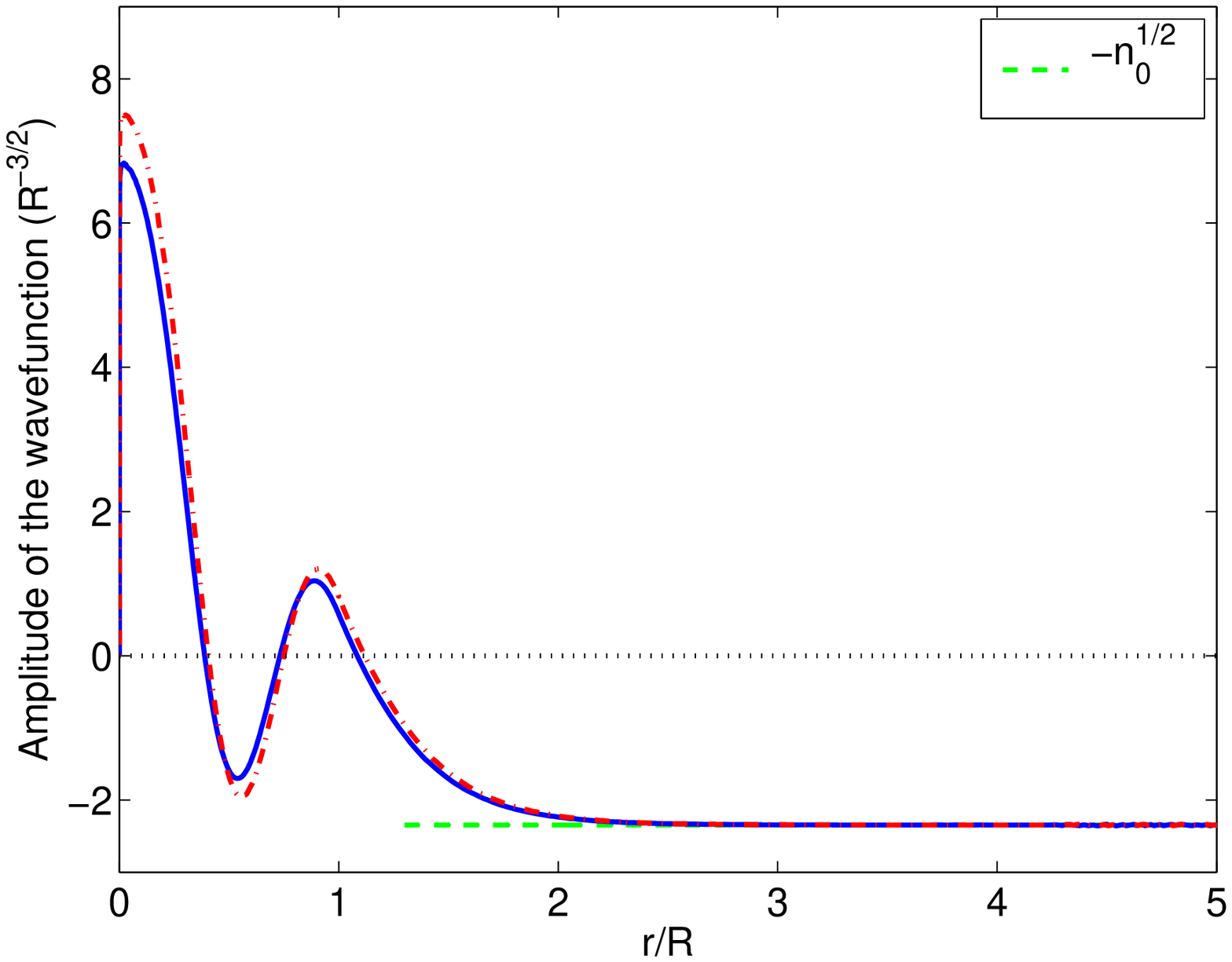}

\caption{\label{cap:merging}(Color online) Two solutions of the Gross-Pitaevskii equation 
  for the attractive square well potential.  $k_0R$ and $U_0$ are the
  same as in Fig.\ \ref{cap:spirals}. For the upper panel, the
  chemical potential is $0.45$, as in Fig.\ \ref{cap:spirals}a, while
  for the lower one it is $2.5$, just below the value at which the
  solutions merge.  The solutions both have three nodes, and are the
  first to merge as the chemical potential increases.}
\end{figure}

In Figs.\ \ref{cap:spirals} and \ref{cap:merging} we show how,
increasing the external density, the solutions with the highest number
of nodes actually merge. For densities higher than this critical
value, the only solutions are ones with a smaller number of nodes.
 
Despite its short-range character, the model given above captures the
main features of the solutions of the GP equation for the atom-ion
potential, which is long-ranged.  We note that the discontinuous
behavior does not occur in a one-dimensional model.

\section{The $r^{-4}$ potential} 

\label{polpot}

We now turn to a more realistic potential with the same $r^{-4}$ behavior
as the actual atom-ion interaction at large distances.  For definiteness, we consider parameters
appropriate for a $^{87}$Rb condensate, and we take $a_{\rm aa}=100\,a_0$.
At large distances, we take the atom-ion potential to be given by Eq.\
(\ref{pot}) with $\tilde \alpha=320a_0^3$.  The wave functions are
sensitive to the short-range behavior of the potential, but we may
obtain illustrative results by cutting the $1/r^4$ potential off by a
hard core of radius $R$. Since
many atoms are bound to the ion, we assume the ion to be static and
set $m_{\rm ai}= m$. The atom-ion scattering length of such potential may
be calculated in the WKB approximation, and is given by
\cite{GribFlam}
\begin{equation}
a_\mathrm{ai}=\beta_{4}\cot\left[\frac{\beta_{4}}{R}\right].
\label{eq: a(ai)}
\end{equation}
The number of bound states allowed by the potential can be estimated by increasing the potential strength from 0 to its actual value. A bound state appears each time the scattering length diverges, and therefore the number of bound states is given by 
\beq
\nu_{\rm S}={\rm Int}\left(\frac{\beta_4}{\pi R}\right),
\label{bs}
\eeq
where ${\rm Int}(x)$ denotes the integer part of $x$.
To model actual atom-ion potentials, a physically reasonable value of
$R$ would be $\sim 10a_0$.  However, the properties of the wave
function of most importance here are those at relatively large
distances, $r\ga 10^3 a_0$, so we take $R=300a_0$, since this should
give us the correct physical behavior for the distances of
interest. We do not expect the qualitative behavior of the wave
function to depend on $R$, even though quantities like the scattering
length do, and we have verified this numerically.

\begin{figure}
\includegraphics[width=\columnwidth,clip=]{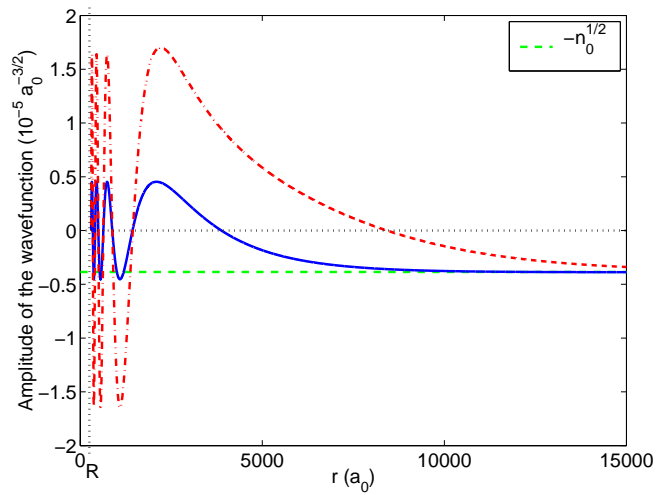}     
\caption{\label{cap: 2 solutions}
(Color online) Condensate wave functions for the two uppermost states in the $1/r^4$
potential with the parameters given in the text for
$n_0=10^{14}$cm$^{-3}$. Both states have seven nodes, but the
resolution of the figure is inadequate to exhibit the rapid
oscillations for $r$ close to $R$. The state that, in the dilute limit, becomes 
the zero-energy solution of the Schr\"odinger equation is given by the solid
line.}
\end{figure}

We now describe numerical solutions of the GP equation that approach a
constant density $n_{0}$ far from the ion. Just as for the finite-range 
potential considered in the previous section, there is generally
more than one solution for a given value of the chemical potential,
and for small external densities one expects $2\nu_{\rm S} +1$. In
Fig.\ \ref{cap: 2 solutions} we show the wave functions
corresponding to the two states with the highest number of nodes,
namely seven for the parameters chosen, in agreement with the
quasiclassical result (\ref{bs}).  The free energy, for a given
condensate density $n_0$, is highest for the states with the highest
number of nodes, and decreases as the number of nodes decreases.

In the absence of inelastic processes, we expect only the
uppermost state of the ionic potential to play an important role in
the capture process, since it is the only one with an appreciable
overlap with the continuum wave function representing the unbound atoms
\cite{Cote'}.

\begin{figure}
\includegraphics[width=\columnwidth,clip=]{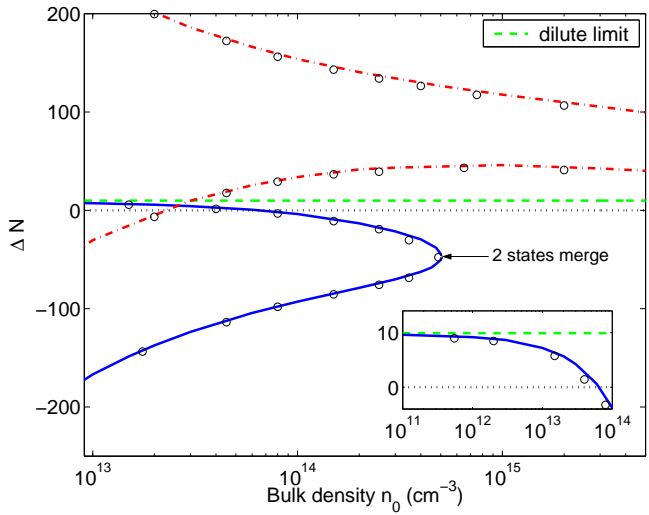}
\caption{\label{cap:Surplus-of-atoms}(Color online) Excess number of atoms around
  a single ion as a function of the bulk density.  The dashed line is
  the dilute limit appropriate for a fixed ion, $\Delta N=-a_{\rm
    ai}/2a_{\rm aa}$ ($R=300\, a_0$ gives $a_{\rm ai}\approx -1980a_0$
  for an infinitely massive ion). Results are shown for the four
  uppermost levels for this potential (i.e.\ the two with 7 nodes and
  the two with 6 nodes, indicated respectively by the solid and
  dot-dashed lines). The lines are obtained from Eq.\ (\ref{DeltaN}),
  the circles from Eq.\ (\ref{DeltaN-FreeEnergy}). The inset exhibits
  the behavior at lower densities.}
\end{figure}

The excess number of atoms is given in terms of the atomic density distribution by  
Eq.\ (\ref{DeltaN}) or, 
alternatively,  from the free energy $F=E-\mu N$ by Eq.\ (\ref{DeltaN-FreeEnergy}).
In Fig. \ref{cap:Surplus-of-atoms} we show results obtained from our
numerical simulations by both methods. In the limit of very low
condensate density we get values for $\Delta N$ in accord with the
thermodynamic arguments in Sec. \ref{sec:thermo}. The consistency of
the two methods of calculation has been confirmed for core radii that
give scattering lengths in the range $\left|a_{\rm ai}\right|<5000\,
a_{0}$.

\begin{figure}
\includegraphics[width=\columnwidth,clip=]{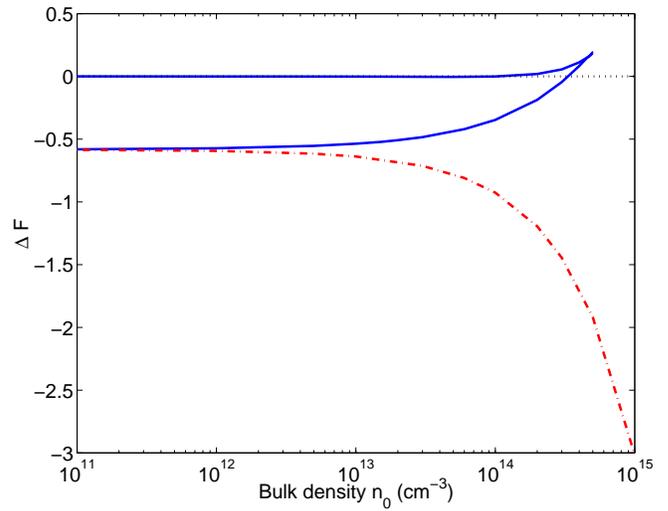}
\caption{\label{cap:DeltaFreeEn}
  (Color online) Difference in free energy for the states given in the
  previous figure: the solid lines are for the two states with 7 nodes
  and the dot-dashed line for one of the states with 6 nodes. The free
  energy is measured in units of 10$^{-5}\hbar^2/ma_0^2$. The other
  state with 6 nodes lies much lower, at around $\Delta F \approx
  -3\cdot10^{-4} \hbar^2/ma_0^2$.}
\end{figure}

The figure shows that the excess numbers of atoms for two states with
the same number of nodes become equal at the density above which the
solutions no longer exist.  This is to be expected, since the
solutions become identical at this point.  In Appendix A we use
quasiclassical arguments to estimate the density at which solutions
merge and disappear, and these are in good agreement with the
numerical results.

In the detailed calculations described so far we have focused
attention on states with close to the maximum number of nodes.  In
particular, in the low-density limit and in the absence of inelastic
processes that can cause the system to relax, one would expect the
state of the condensate to be the one that close to the ion resembles
the zero-energy solution of the Schr\"odinger equation.  However,
three-body processes can relax the system, thereby populating states
with lower numbers of nodes.  To calculate properties of such a
system, one could start with a many-particle wave function of the
Hartree-Fock type in which more than one single-particle state is
occupied, and solve the Hartree-Fock equations.  This is, however,
beyond the scope of this paper because the density of atoms rises to
values sufficiently high that the dilute gas approximation for the
interaction energy employed in the GP approach fails at relatively
large distances from the ion.  To see this, we note that the density
of atoms far from the ion will be given by the Thomas--Fermi
approximation, Eq.\ (\ref{TFsolution}).  The dilute gas approximation
is valid provided $n|a_{\rm aa}|^3\ll 1 $.  This condition becomes
\beq n|a_{\rm aa}|^3 \approx \left|\frac{V(r)}{U_0}a_{\rm aa}^3\right|
=\frac{\beta_4^2 a_{\rm aa}^2}{8 \pi r^4} \ll 1 \eeq or \beq r\gg
(\beta_4 |a_{\rm aa}|)^{1/2}/2, \eeq which for rubidium ($a_{\rm
  aa}\approx 100 a_0$) implies that the GP equation is valid only for
$r \gg 400 a_0 $ for such states.

\section{Conclusions and discussion}
\label{concl}

In this paper we have investigated solutions of the Gross-Pitaevskii
equation for a Bose--Einstein condensate in the presence of a positive
ion. We find that for low condensate densities, there are $2\nu_{\rm
S}+1$ solutions for a given condensate density, where $\nu_{\rm S}$ is
the number of bound states of the Schr\"odinger equation.  With increasing condensate density, pairs of states become
degenerate and disappear, and the state of the system must change
discontinuously.  An interesting challenge is to find experimental
evidence for such a behavior.

We have calculated the excess number of atoms around an ion, and find
that for the state that resembles the zero-energy solution of the
Schr\"odinger equation it can be either positive or negative,
depending on the sign of the atom--ion phase shift, and a typical
magnitude is of order $\sim 10^2$ . The spatial size of the density
disturbance around an ion is set by $\beta_4 \sim 1$ $\mu$m.  Our
estimates indicate that the Gross-Pitaevskii equation should give a
reliable first approximation for the wave function of such states. For
states with fewer nodes, the density of atoms may reach values high
enough that the GP equation fails.  

There are many outstanding problems.  In most of the calculations we
have assumed that the state of interest is that with the maximum
possible number of nodes.  More study is needed of inelastic processes
that will cause atoms to relax to lower states \cite{Cote'}.
Experimental studies will be valuable in providing guidance for future
work.

\subsubsection*{Acknowledgments}

We are grateful to Halvor Nilsen for valuable discussions.  We also
thank Andrew Jackson and Alexander Lande for helpful insights, and
Marco Anderlini and Vasili Kharchenko for useful correspondence.
Part of this work was done while one of the authors (CJP)
participated in the program on Quantum Gases at the Kavli Institute
for Theoretical Physics at the University of California, Santa
Barbara. This research was supported in part by the National Science
Foundation under Grant No. PHY99-07949.

\appendix*
\section{Validity of approximations}

Here we examine the conditions under which it is a good approximation
to replace the effective atom-atom interaction by the standard expression
(\ref{U}).  Later we shall estimate the density below which the dilute
gas result (\ref{eq: deficit in the dilute limit1}) would be expected
to hold.  For nonzero wave numbers $k$, the quantity that enters the
expression for the energy shift is $-\delta/k$, where $\delta$ is the $s$-wave phase shift, 
rather than $a$
\cite{mahan}.  Since a typical energy scale for changes in $\delta/k$
is set by $\hbar^2/2m\beta_6^2$, while the potential depth is given by
$\tilde{\alpha}e_0^2/2r^4$, we expect that the scattering length approximation will
fail when
\beq\frac{\tilde\alpha e_0^2}{2r^4}\gg\frac{\hbar^2}{2m\beta_6^2}\ \ \ \mathrm{or}\ \ \ r^4\ll\beta_4^2\beta_6^2.\eeq
The Gross-Pitaevskii approach should therefore be valid if the phase
shift due to the region where $r\ll\bar{r}=(\beta_4\beta_6)^{1/2}$ is
negligible. To estimate this phase shift, we make a semiclassical
approximation to the GP equation.  This gives for the total accumulated
phase out to a distance $r$ 
\begin{equation}
\Phi(r) \approx \int^r dr'\sqrt{\frac {2m}{\hbar^2}[\mu-V(r')-n(r')U_0]}.
\label{Phi}
\end{equation}
Deep in the ionic potential, the wave function is given to a good
approximation by the semiclassical result, which has an amplitude
\beq
\psi \propto (r p_{\rm cl}^{1/2})^{-1},
\eeq
where $p_{\rm cl}(r) = [2mV(r)]^{1/2}$ is the classical
momentum of a particle of zero total energy in the presence of the
potential.  For the $r^{-4}$ potential, the amplitude of the wave
function is therefore independent of $r$, and therefore we may replace
the mean-field energy to a first approximation by a constant
$\tilde{n} U_0$, where $\tilde n$ is independent of $r$.  Expanding
expression (\ref{Phi}) in the deviation $n_0-\tilde n$ we find

\begin{equation}\Phi(r)\approx
\Phi_0(r)+(n_0-\tilde{n})\frac{2 m U_0}{\hbar^2}\int_0^{r}dr'\frac {1} {2\sqrt{-2mV(r')/\hbar^2}}.
\end{equation}
Due to the mean-field interaction, the accumulated phase out to a
distance $r \sim {\bar r}= (\beta_4 \beta_6)^{1/2}$ is therefore changed by an
amount
\beq
\delta \Phi(\bar r) \approx    (n_0-\tilde n)a_{\rm aa}  (\beta_4 \beta_6^3)^{1/2}. 
\eeq
If we take the interior density to be of the same order of magnitude as that far from the ion, one finds
\beq
\delta \Phi(\bar r) \sim \frac{\beta_4^{1/2} \beta_6^{3/2}}{\xi^2}, 
\eeq
where the healing length is defined in Eq.\ (\ref{healing}).  Since
under experimental conditions the healing length is typically
comparable to $\beta_4$, while $\beta_6$ is two orders of magnitude
smaller, this shows that the region close to the ion where the
Gross-Pitaevskii equation fails is likely to be unimportant.

On the basis of the above calculation, we may also estimate the
density below which the low-density result (\ref{eq: deficit in the
dilute limit1}) is valid.  Using the approximations above, we find
that the total accumulated phase out to a distance $\sim \beta_4$,
where the semiclassical treatment fails, is of order
\beq
\delta \Phi( \beta_4 ) \sim \frac{\beta_4^2}{\xi^2}. 
\eeq
 This indicates that changes to the accumulated phase can be
 significant under typical experimental conditions.

%Data for Rb: \beta_4: 380 nm, \beta_6: 8.76 nm, \sqrt{\beta_4\beta_6}=58 nm


\begin{thebibliography}{1}

  
\bibitem{Pisa}D.\ Ciampini, M.\ Anderlini, J.\ H.\ M\"uller, F.\ Fuso,
  O.\ Morsch, J.\ W.\ Thomsen, and E.\ Arimondo, Phys.\ Rev.\ A
  \textbf{66}, 043409 (2002).

\bibitem{Cote'}R.\ C\^ot\'e, V.\ Kharchenko, and M.\ D.\ Lukin, Phys.\ 
  Rev.\ Lett.\ \textbf{89}, 093001 (2002).

\bibitem{bbp}J.\ Bardeen, G.\ Baym, and D.\ Pines, Phys.\ Rev.\ Lett.\ 
  \textbf{17}, 372 (1966); Phys.\ Rev.\ \textbf{156}, 207 (1967).
  
\bibitem{GribFlam}G.\ F.\ Gribakin and V.\ V.\ Flambaum, Phys.\ Rev.\ 
  A \textbf{48}, 546 (1993).

  
\bibitem{mahan} F.\ G.\ Fumi, Phil. Mag. {\bf 46}, 1007 (1955); G.\ 
  D.\ Mahan, \textit{Many-particle physics}, Second edition (Plenum,
  New York, 1990), p. 253.

\end{thebibliography}
\end{document}